\pdfoutput=1

\documentclass[aps,prd,reprint,showpacs,showkeys]{revtex4-1}
\usepackage{amsmath}
\usepackage{bm}
\usepackage{graphicx}
\usepackage[pdftex,colorlinks,hyperindex,plainpages=false,bookmarksopen,bookmarksnumbered,pdfusetitle,allcolors={blue}]{hyperref}
\usepackage{float}
\usepackage{mathrsfs}
\usepackage{subcaption}
\usepackage{caption}
\begin{document}

\title{Scattering theory of transport through disordered magnets}

\author{Martin F.~Jakobsen, Alireza Qaiumzadeh, Arne Brataas}
%\email{martin.f.jakobsen@ntnu.no}
\affiliation{Center for Quantum Spintronics, Department of Physics, Norwegian University of Science and Technology, NO-7491 Trondheim, Norway}

%\author{A. Qaiumzadeh}
%\email{alireza.qaiumzadeh@ntnu.no}
%\affiliation{Department of Physics, Norwegian University of Science and Technology, NO-7491 Trondheim, Norway}
%\author{A. Brataas}
%\email{arne.brataas@ntnu.no}
%\affiliation{Department of Physics, Norwegian University of Science and Technology, 7491 Trondheim, Norway}

\begin{abstract}
We present a scattering theory of transport through noncollinear disordered magnetic insulators. For concreteness, we study and compare the random field model (RFM) and the random anisotropy model (RAM). The RFM and RAM are used to model random spin disorder systems and amorphous materials, respectively. We utilize the Landauer-B{\"u}ttiker formalism to compute the transmission probability and spin conductance of one-dimensional disordered spin chains. The RFM and the RAM both exhibit Anderson localization, which means that the transmission probability and spin conductance decay exponentially with the system length. We define two localization lengths based on the transmission probability and the spin conductance, respectively. Next, we numerically determine the relationship between the localization lengths and the strength of the disorder. In the limit of weak disorder, we find that the localization lengths obey power laws and determine the critical exponents. Our results are expressed via the universal exchange length and are therefore expected to be general.
\end{abstract}

%\pacs{???}
%\keywords{???}
\maketitle

%%%%%%%%%%%%%%%%%%%%%%%%%%%%%%%%%%%%%%%%%%%%%%%%%%%

\section{Introduction}
\label{sec:intro}
In magnonics~\cite{Kruglyak_2010,Chumak2015,Serga_2010,KRUGLYAK2006191,PhysRevB.97.020402,PhysRevLett.112.147204,doi:10.1002/adma.200900809,PhysRevLett.88.117601}, the primary focus has recently been on the propagation of spin waves through various types of magnetic insulators. A particular emphasis has been on ordered systems, such as (anti)ferromagnets, and ferrimagnets. An advantage is that the spin current may suffer less Joule heating compared to electric currents, making insulator-magnonics applications potentially much more energy efficient~\cite{Serga_2010,Bauer2012}. Numerous successful experiments have generated and manipulated spin currents using the spin Hall effect and the inverse spin Hall effect~\cite{RevModPhys.87.1213}. A common experimental setup consists of sandwiching a magnetic insulator between two conductors and using the spin Hall effect to generate a spin current in the left conductor that propagates through the magnetic insulator and into the right conductor. The spin current in the right conductor is converted into a charge current via the inverse spin-Hall effect. This provides a useful method to measure the spin current and infer the spin-transport properties of the magnetic insulator~\cite{PhysRev.95.1154,DYAKONOV1971459,PhysRevLett.85.393,Murakami1348,PhysRevLett.92.126603,Jungwirth2012,Kajiwara2010,doi:10.1063/1.2199473,Valenzuela2006,PhysRevLett.98.156601,PhysRevLett.99.266603,Seki2008,PhysRevB.78.014413,doi:10.1063/1.3327809,Cornelissen2015}.

A class of materials that has recently attracted attention in the spintronics community is disordered magnetic insulators~\cite{disorderHydrodynamics,spintransfertorque,disorderedSpiral,PhysRevB.97.184423,symmetryMacroscopicDynamics,Buczek_2018}. Notably, a recent experiment claimed that a spin current flowing through a sample of amorphous yttrium-iron-garnet could travel tens of micrometers~\cite{longDistanceSpinTransport}. This distance is comparable to the spin current propagation length in a crystalline (anti)ferromagnet~\cite{ArneLongDistance2018,Cornelissen2015}. More generally, it is crucial to study disordered magnetic materials because almost all materials contain some degree of disorder, which will affect the functional properties of magnonic devices. When the disorder is sufficiently strong, the eigenstates become trapped in a finite spatial region, completely suppressing the transport properties. This phenomenon is known as Anderson localization, and the first discussion of this phenomenon in magnetic systems began in the 1960s~\cite{ClassicsHydrodynamicSpinGlass,spinResonanceGlass, andersonGlass, ANDERSON1970549, Korenblit1978,macroscopicNonCollinear,excitationsSpinGlasses,Weber1969,Barnes1981}. Furthermore, it has been shown that even with a small onset of disorder, the transport properties change from conductive to diffusive~\cite{PhysRevB.97.184423,PhysRevB.97.020407}, which has important consequences for magnonics applications in low dimensions.

The common sources of quenched disorder in magnetic insulators are $i)$ randomness due to anisotropies, local fields, and amorphous structure and $ii)$ frustration due to competing long-range exchange interactions. In this paper we focus on magnetic insulators with quenched disorder due to $i)$. Two models with these properties are the random field model (RFM) and the random anisotropy model (RAM), where the disorder is caused by the competition between the exchange interaction and the coupling to local random fields and anisotropies respectively. The RFM and RAM is used to model quenched spin disorder and amorphous magnets respectively~\cite{dedominics,10.1007/1-4020-2965-9_2,doi:10.1063/1.324881,PhysRevLett.31.160,PhysRevB.95.125129,PhysRevE.86.021131}. Experimental realizations of such systems are plentiful~\cite{TheoryOfSpinGlasses,PhysRevLett.35.1792,MATTIS1976421,doi:10.1142/3517,PhysRevLett.50.1946}.

Furthermore, there are two types of RFM/RAM spin models. The first is the Ising model, where the spins are scalars $S_i = \pm 1$ and are randomly pointing either parallel or antiparallel to each other in the ground state~\cite{Smith2016,BELANGER1991272,PhysRevLett.116.227201,PhysRevLett.93.267008,PhysRevLett.53.1747,PhysRevE.73.016109,Fishman_1979,PhysRevB.29.505}. The second type is the Heisenberg model, where the spins are vectors $\mathbf{S}_i$ that in the ground state are pointing noncollinearly in random directions~\cite{PhysRevB.91.081103,PhysRevB.82.174411,De_Luca_2013,PhysRevB.90.064201,PhysRevB.77.064426,PhysRevB.93.224205}.

Because the ground state in the RFM/RAM Ising model is relatively simple, it can often be studied efficiently with analytical methods. For example, one can either solve the equations of motion by a transfer matrix approach paralleling Anderson's celebrated work on disordered fermionic systems~\cite{Anderson:Localization,PhysRevLett.48.699} or one can use field-theory methods, particularly the replica trick, replica symmetry breaking, and mean-field theory~\cite{PhysRevLett.41.1068,PhysRevLett.43.1754,PhysRevLett.50.1946}. Although the RFM/RAM Ising models are analytically accessible, they are only simplified idealizations of a real disordered magnet where the spins are noncollinear. In this work, we wish to focus on systems with noncollinear spins that are harder to describe analytically but exhibit more realistic spin-wave dynamics.

Disordered magnetic insulators with a noncollinear ground state are a notoriously difficult system to describe. Due to their complexity, it is often useful to study the classical spin waves of the system. Our work is related to a recent study~\cite{EveresNowak:disorder2015, PhysRevB.97.184423} in which the micromagnetic Landau-Lifshitz-Gilbert  (LLG) equation was solved using a quasimonochromatic Gaussian wave packet as the initial condition. They found that the width of the wave packet increases in time until it saturates around the localization length of the system; a hallmark of Anderson localization. In systems that exhibit Anderson localization, the localization length decreases as the system becomes more disordered. However, the exact relationship between the localization length and the strength of disorder is far from being well established in noncollinear disordered magnetic insulators. In this work, we attempt to shed some light on these issues.

The localization effect in spin models depends on the dimensionality of the system, similar to disordered fermionic systems \cite{PhysRevLett.42.673}. For fermionic systems in one dimension, there is Anderson localization; in two dimensions, the effect remains present but much weaker, while in three dimensions, there is the possibility of both a localized and a delocalized phase. The same observations have been established for disordered magnets \cite{TheoryOfSpinGlasses,cuspExperiment}. We focus on one-dimensional spin chains. With more computational time, this method can also be applied to two- and three-dimensional systems.

%%%%%%%%%%%%%%%%%%%%%%%%%%%%%%%%%%%%%%%%%%%%%%%%%%%%%

The numerical method that we develop is based on the Landauer-B{\"u}ttiker formalism ~\cite{Landauer,RevModPhys.69.731}, which has proven to be extremely useful in studying the transport properties of electronic systems. To the best of our knowledge, such a method has not previously been applied to the RFM/RAM Heisenberg model. In this paper, we investigate the effect of Anderson localization on the spin-wave transport properties of a disordered magnetic insulator. To this end, we first determine the relationship between the system size and the transmission probability for different strengths of disorder and then calculate the spin conductance. With this knowledge, we can investigate how the localization length of the system scales with the strength of the disorder. In particular, we calculate and compare the critical exponents of the RFM and the RAM. These quantities provide us with direct insights into how the transport properties of the spin waves are affected by the localization phenomenon that is present due to quenched disorder.

We hope that this theoretical investigation may inspire an experimental investigation into the transport properties of disordered magnetic nanowires~\cite{doi:10.1116/1.4929897,PhysRevB.22.3344,PhysRevB.25.2681,doi:10.1063/1.333643,KANEYOSHI201534}. In particular, it would be interesting to compare the experimental relationship between the localization length and strength of disorder to the critical exponents that we determine in this work.

The paper is organized as follows. In Sec.~\ref{sec:current},
we introduce the RFM and the RAM Hamiltonians and discuss its ground state. In section \ref{sec:SCATTERING THEORY} we find the linearized equations of motion, and derive expressions for the spin current and the spin conductance. Section \ref{sec:results}. contains our numerical calculations of the scattering properties of the system. In Sec. \ref{sec:results} we summarize our results.
%%%%%%%%%%%%%%%%%%%%%%%%%%%%%%%%%%%%%%%%%%%%%%%%%%%
\section{THEORETICAL MODEL}
In this section we carefully introduce the model we are interested in studying. We start by presenting the Hamiltonians for the RFM and the RAM, and introduce the geometry. We conclude this section by presenting a method to calculate the classical metastable states.
\label{sec:current}
\subsection{Hamiltonian}
To investigate the transport properties of one-dimensional disordered noncollinear spin chains, we use the Hamiltonian
\begin{equation}
    H^{\kappa} = - J \sum_i \mathbf{S}_i \cdot \mathbf{S}_{i+1}- K \sum_i \left(\mathbf{S}_i \cdot \mathbf{n}_i\right)^{\kappa + 1},
    \label{Eq:SpinHamiltonian}
\end{equation}
where $\kappa = 0$ and $\kappa = 1$ represent the RFM and the RAM, respectively.

The dimensionless spins $\mathbf{S}_i$ are attached to a one-dimensional lattice with lattice spacing $d$. The exchange interaction with $J>0$ attempts to align the spins. The terms proportional to $K$ encapsulate the quenched disorder of the system, and we choose $K> 0$ without loss of generality. Each spin $\mathbf{S}_i$ is coupled to a local random vector $\mathbf{n}_i$. The competition between the exchange and the random interactions in Eq.~\eqref{Eq:SpinHamiltonian} results in a noncollinear disordered ground state. We use the parameter $K/J$ to characterize the strength of disorder.
%%%%%%%%%%%%%%%%%%%%%%%%%%%%%%%%%%%%%%%%%%%%%%%%%%%%%%%%
\subsection{Geometry}
\begin{figure}[h]
\centering
\includegraphics[width=\columnwidth]{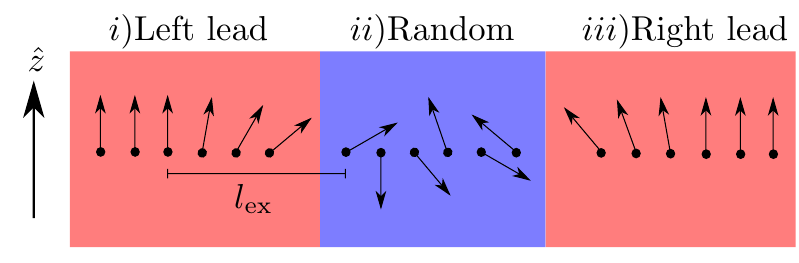}
\caption{A disordered magnet (blue) is sandwiched between two ferromagnets (red). In regions $i)$ and $iii)$, $\mathbf{n}_i = \hat{z}$, while in region $ii)$ $\mathbf{n}_i$, is uniformly distributed on the unit sphere. Consequently, the spins in region $ii)$ point in random directions, while the spins deep inside regions $i)$ and $iii)$ point in the $z$-direction. The spins close to the two interfaces rotate similar to the spins in a domain wall. The length of the domain-wall-like region is illustrated and given by the exchange length $l_{\mathrm{ex}} = \sqrt{J/K}d$. }
\label{fig:Setup_manuscript4}
\end{figure}

We consider a one-dimensional chain with $N$ lattice sites. The chain is split into three regions that we call $i)$ the left lead, $ii)$ the random region, and $iii)$ the right lead; see Fig. \ref{fig:Setup_manuscript4}.
\\
\noindent
In regions $i)$ and $iii)$, we let the number of spins be equal to $N_\mathrm{L}$ and $N_\mathrm{R}$, respectively. In addition, we let $\mathbf{n}_i$ point in the $\hat{z}$-direction. In region $ii)$, we let the number of spins be equal to $N_{\mathrm{rand}}$ and $\mathbf{n}_i$ to point in some random direction uniformly distributed on the unit sphere. Note that far away from the random region (deep inside of the leads), the spins point in the $\hat{z}$-direction, while in the random region, the spins are oriented randomly. In the regions close to the interface, the spins are rotating in a domain-wall-like fashion. The length of this domain-wall region is given by the exchange length $l_{\mathrm{ex}} = \sqrt{J/K}d$.

The scattering problem that we are interested in studying can now be realized by exciting coherent spin waves in the left lead propagating towards the random region. As the spin wave approaches the random region, it will be scattered either back into the left lead (reflection) or into the right lead (transmission). We assume semi-infinite leads such that $N_\mathrm{L}$ and $N_\mathrm{R} \xrightarrow{}\infty.$
%%%%%%%%%%%%%%%%%%%%%%%%%%%%%%%%%%%%%%%%%%%%%%%%%%%%%%%%
\subsection{The ground state}
Determining the true ground state of a disordered magnet (collinear or noncollinear) is a very challenging problem. The primary reason is that the randomness results in free energy with many nearly degenerate minima, separated by high energy barriers. The problem of determining the exact ground state of disordered systems is its own research field, and we do not wish to address that problem here~\cite{Edwards_1980,PhysRevB.70.064412,PhysRevLett.120.087201,Castellani_2005,doi:10.1142/0271,Bray_1981,PhysRevB.24.6579,PhysRevE.60.5244,PhysRevE.73.036110,PhysRevE.76.051107,PhysRevE.77.041114,Heuer_2008}. However, due to the high energy barriers, the probability of tunneling between different metastable states is small. Hence, in an experiment, the disordered magnet becomes trapped in a state that may differ from the exact ground state when the system is cooled down, depending on the history. Thus, in this paper, we study the transport properties of disordered magnets around classical metastable states.

We can find a classical metastable state of the system by treating the spins as classical vectors obeying the LLG equation of the form
\begin{equation}
    \frac{d \mathbf{S}_i}{dt} = -\gamma \mathbf{S}_i \times \mathbf{\mathcal{H}}_i^{\kappa} - \lambda \mathbf{S}_i \times \left(\mathbf{S}_i \times \mathbf{\mathbf{\mathcal{H}}}_i^{\kappa}\right).
    \label{Eq:LLG}
\end{equation}
Here, the first term with $\gamma > 0$ describes the spin $\mathbf{S}_i$ precessing around its instantaneous effective field $\mathbf{\mathcal{H}}_i^{\kappa} = -\delta H^{\kappa} / \delta \mathbf{S}_i$, while the second term describes the damping towards the direction of the instantaneous effective field. The metastable state is then obtained by specifying some arbitrary initial configuration and allowing the spins to evolve according to this equation for sufficiently long times $t\xrightarrow{} \infty$.
%%%%%%%%%%%%%%%%%%%%%%%%%%%%%%%%%%%%%%%%%%%%%%%%%%%%%%%%
\section{SCATTERING THEORY}
\label{sec:SCATTERING THEORY}
In this section we outline the theoretical approach that we will use to determine the transport properties of the RFM and the RAM. We start by determining the linearized equations of motion, and formulate the scattering problem. Finally, we derive the expressions for the spin current and spin conductance in the linear response regime.
\subsection{Hamiltonian in terms of spin-wave operators}
To study the transport properties of the system, we can perform a Holstein-Primakoff expansion around one of the metastable states. Let us at each site $i$ define a local coordinate system $\{\hat{e}_x(i),\hat{e}_y(i),\hat{e}_z(i)\}$ such that $\hat{e}_z(i)$ is parallel to the spin at site $i$ in the ground state. The spin operator in a low-lying excited state can then be written as
\begin{equation}
    \mathbf{S}_i = \hat{e}_z(i) S_{i}^z + \hat{e}_x(i) S_{i}^x + \hat{e}_y(i) S_{i}^y.
    \label{Eq:SpinDecomposition}
\end{equation}
We perform a Holstein-Primakoff transformation of the form
\begin{subequations}
\begin{align}
& S_{i}^x  \approx \sqrt{\frac{S}{2}}\left(a_i+a^\dagger_i\right),\\
& S_{i}^y  \approx -i\sqrt{\frac{S}{2}}\left(a_i-a_i^\dagger\right),\\
& S_{i}^z = S-a^\dagger_ia_i.
\end{align}
\label{Eq:HPSPINCOMPONENTS}
\end{subequations}
In Eq.~\eqref{Eq:HPSPINCOMPONENTS}, we have only included the lowest-order terms because we are not interested in studying the interactions between the spin waves. If we substitute Eqs. \eqref{Eq:SpinDecomposition} and \eqref{Eq:HPSPINCOMPONENTS}  into Eq. \eqref{Eq:SpinHamiltonian} and introduce the notations $\hat{e}_{\pm}(i) = \hat{e}_x(i) \pm i \hat{e}_y(i)$ and $n_i^{\pm} = n^x_i \pm i n_i^y$, we obtain a Hamiltonian of the form
\begin{equation}
    H^{\kappa} = \sum_{ij} A_{ij}^{\kappa}a_i^{\dagger}a_j + B_{ij}^{\kappa} a_i a_{j} + h.c,
    \label{Eq:QuadraticHamiltonian}
\end{equation}
where
\begin{equation}
    \begin{split}
        & A_{ij}^{\kappa} = \delta_{i,j} \Big\{ JS \hat{e}_z(i) \hat{e}_z(i+1) + \frac{1}{2}K n_i^z \\
        &\qquad+\kappa \big[ K S (n_i^z)^2 - \frac{1}{2}KS n_i^- n_i^+ - \frac{1}{2} K n_i^z \big]\Big\}\\
        &\qquad- \frac{JS}{2} \delta_{i,j+1} \hat{e}_-(i) \hat{e}_+(j),\\
        &B_{ij}^{\kappa} = - \kappa \frac{KS}{2}(n_i^-)^2\delta_{i,j} - \frac{JS}{2}\hat{e}_-(i) \hat{e}_-(i+1) \delta_{i,j+1}.
    \end{split}
\end{equation}
In the following, we will study the spin waves associated with the Hamiltonian of Eq.~\eqref{Eq:QuadraticHamiltonian}.

%%%%%%%%%%%%%%%%%%%%%%%%%%%%%%%%%%%%%%%%%%%%%%%%%%%%%%%%
\subsection{Equations of motion}
The equations of motion for the spin-wave operators can now be calculated from the Heisenberg equation
\begin{equation}
    \frac{d}{dt} a^{\pm}_i = \frac{i}{\hbar} \left[H^{\kappa},a^{\pm}_i\right].
\end{equation}
 For clarity, we reinstate the spin operators $\{S_i^x,S_i^y\}$ using Eq. \eqref{Eq:HPSPINCOMPONENTS} and cast the equation of motion in the form
\begin{widetext}
\begin{equation}
    \begin{split}
        \hbar\frac{d S_j^x}{dt} &= JS \Big\{ \hat{e}_z(j) \big[ \hat{e}_z(j+1) + \hat{e}_z(j-1)\big] S_j^y \\
        &-\hat{e}_y(j-1) \hat{e}_y(j) S_{j-1}^y - \hat{e}_x(j-1) \hat{e}_y(j) S_{j-1}^x -\hat{e}_y(j)\hat{e}_y(j+1) S_{j+1}^y - \hat{e}_y(j) \hat{e}_x(j+1) S_{j+1}^x\Big\}\\
        &+  K n_j^z S_j^y+\kappa \Big\{ 2 K S (n_j^z)^2 S_j^y - 2 K S (n_j^y)^2 S_j^y - 2 K S n_j^x n_j^y S_j^x - K n_j^z S_j^y \Big\},
    \end{split}
    \label{Eq:LLGx}
\end{equation}
\begin{equation}
    \begin{split}
        \hbar\frac{d S_j^y}{dt} &= JS \Big\{ -\hat{e}_z(j) \big[ \hat{e}_z(j+1) + \hat{e}_z(j-1)\big] S_j^x\\
        &+\hat{e}_x(j-1) \hat{e}_x(j) S_{j-1}^x + \hat{e}_y(j-1) \hat{e}_x(j) S_{j-1}^y +\hat{e}_x(j)\hat{e}_x(j+1) S_{j+1}^x + \hat{e}_x(j) \hat{e}_y(j+1) S_{j+1}^y\Big\} \\
        &- K n_j^z S_j^x - \kappa \Big\{ 2 K S (n_j^z)^2 S_j^x - 2 K S (n_j^x)^2 S_j^y - 2 K S n_j^x n_j^y S_j^y + K n_j^z S_j^y \Big\}.
    \end{split}
    \label{Eq:LLGy}
\end{equation}
\end{widetext}
Eqs. \eqref{Eq:LLGx} and \eqref{Eq:LLGy} are identical to the linearized classical Landau-Lifshitz equations expressed in the local coordinate system $\{\hat{e}_x(j),\hat{e}_y(j),\hat{e}_z(j)\}$. Since we are only interested in studying how the intrinsic disorder affects the transport properties of the system, we have not included a Gilbert damping term.

%%%%%%%%%%%%%%%%%%%%%%%%%%%%%%%%%%%%%%%%%%%%%%%%%%%%%%%%
\subsection{Scattering problem and solution ansatz}
Eqs. \eqref{Eq:LLGx} and \eqref{Eq:LLGy} can be solved numerically in the classical regime, where we treat the spin operators as classical vectors. The spin-wave solutions are the normal modes of the system and precess with the same frequency $\omega$. Therefore, we can factorize out the time dependence of the spin operators as $e^{-i \omega t}$.

Deep inside the leads, the spins at neighboring sites are pointing in the $z$-direction; see Fig. \ref{fig:Setup_manuscript4}. This considerably simplifies the equations of motions in the leads
\begin{equation}
\begin{split}
    &-i \hbar \omega S_j^x = JS \left(2 S_j^y - S_{j-1}^y - S_{j+1}^y\right) \\
    &\qquad\qquad \quad+K\left(1+ \kappa \left(2S - 1\right)\right)S_j^y  ,\\
    &-i \hbar \omega S_j^y = JS \left(-2 S_j^x + S_{j-1}^x + S_{j+1}^x\right)\\
    &\qquad\qquad \quad- K\left(1+\kappa\left(2S-1\right)\right)  S_j^x.
    \label{Eq:EOMLEAD}
\end{split}
\end{equation}
The system behaves as a ferromagnet with an external field or intrinsic anisotropy in the $z$-direction. The solutions are therefore circularly polarized plane waves traveling with a fixed wavenumber $q$ and frequency $\omega$. The dispersion relation can be determined by substituting the ansatze $S_x^j = e^{iq jd}$ and $S_y^j = -i e^{iq jd}$ into Eq. \eqref{Eq:EOMLEAD}. The result is
\begin{equation}
    \epsilon = \hbar\omega = 2 JS \left(1-\cos\,qd\right) + K\left(1 + \kappa\left(2S-1\right)\right).
    \label{eq:dispersion_relation}
\end{equation}

Let us now formulate the scattering problem. Deep inside the regions $i)$ and $iii)$ in Fig.~\ref{fig:Setup_manuscript4}, we know that the solution must have the form
\begin{equation}
    \begin{split}
        &S_j^x = e^{iqjd} + r_x e^{-iqjd},\\
        &S_j^y = -i \left(e^{iqjd} + r_y e^{-iqjd}\right),
    \end{split}
    \label{Eq:leftAnsatz}
\end{equation}
and
\begin{equation}
    \begin{split}
        &S_j^x = t_x e^{iqjd},\\
        &S_j^y = -i t_y e^{-iqjd},
    \end{split}
    \label{Eq:rightAnsatz}
\end{equation}
respectively. Inside region $ii)$, we know that the spin components must satisfy Eqs. \eqref{Eq:LLGx} and \eqref{Eq:LLGy}. Using the ansatze as boundary conditions, we have found a finite set of algebraic equations that we can solve numerically to determine the reflection and transmission amplitudes $\{r_x,r_y,t_x,t_y\}$ as functions of $\epsilon$.

%%%%%%%%%%%%%%%%%%%%%%%%%%%%%%%%%%%%%%%%%%%%%%%%%%%%%%%%
\subsection{Spin current and conductance}
Once we know the reflection and transmission amplitude, we can calculate the spin conductance of the disordered magnet utilizing the Landauer-B{\"u}ttiker formalism in the linear response regime. In this section, we derive the expression for spin conductance.

In the leads, the Hamiltonian in Eq. \eqref{Eq:QuadraticHamiltonian} simplifies to
\begin{equation}
\begin{split}
    H^{\kappa} &= \sum_i \left(2 J S + K\left(1 + \kappa\left(2S - 1\right)\right)\right) a_i^{\dagger}a_i\\
    &\quad-JS\left(a_{j} a_{j+1}^{\dagger} + a_j^{\dagger} a_{j+1}\right).
    \end{split}
\end{equation}
From the equation of motion,
\begin{equation}
\begin{split}
     \frac{d}{dt} N_i &= \frac{i}{\hbar} \big[ N_i, H^{\kappa} \big]\\
    &= - i JS \left\{ \left(a_{j+1}^{\dagger} a_j - a_j^{\dagger} a_{j+1}\right) + \left(a_{j-1}^{\dagger} a_j - a_j^{\dagger} a_{j-1}\right) \right\}
\end{split}
\end{equation}
where $N_i = a_i^{\dagger} a_i$ is the number operator, and we can extract the spin current from site $j$ to $j+1$ as
\begin{equation}
    I_{j,j+1} = i J S \left(a_{j+1}^{\dagger} a_j - a_j^{\dagger} a_{j+1} \right).
    \label{Eq:CurrentOperator}
\end{equation}

\begin{figure}[h]
\centering
\includegraphics[width=\columnwidth]{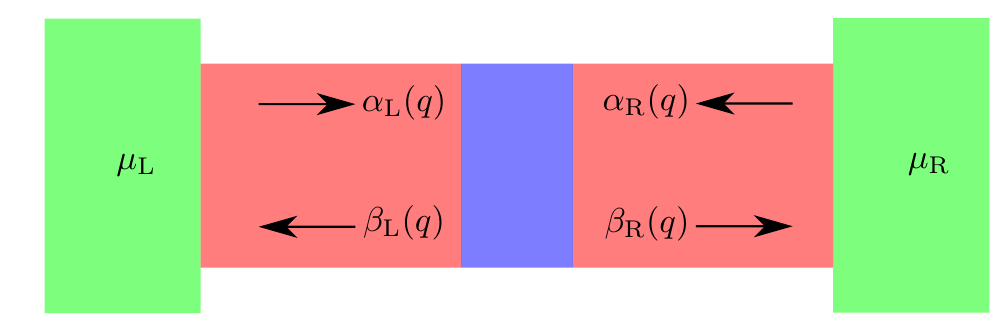}
\caption{
A disordered magnet (blue) is sandwiched between two ferromagnetic leads (red). The leads are connected to two spin reservoirs (green) with spin accumulations $\mu_{\mathrm{L}}$ and $\mu_{\mathrm{R}}$. The reservoirs are in thermodynamic equilibrium such that the magnon population is characterized by the Bose-Einstein distribution. A spin current is induced when there is a nonzero spin bias $\delta \mu = \mu_{\mathrm{L}}-\mu_\mathrm{R}$.
}
\label{Fig:ConductivitySetup_manuscript}
\end{figure}

\pagebreak
Now consider the situation in Fig. \ref{Fig:ConductivitySetup_manuscript}, where two reservoirs in thermodynamic equilibrium are attached to two leads with a scattering region between them. If the spin accumulation in the left reservoir $\mu_{\mathrm{L}}$ is greater than the spin accumulation in the right reservoir $\mu_{\mathrm{R}}$, the spin current in Eq. \eqref{Eq:CurrentOperator} will flow from the left to the right reservoir. We define the operators $\alpha_{\mathrm{L},\mathrm{R}}(q)$ and $\beta_{\mathrm{L},\mathrm{R}}(q)$ injecting and removing magnons with wavenumbers $q$ into the leads, respectively. The relationship between these operators is given by the scattering matrix
\begin{equation}
    \begin{pmatrix} 
\beta_{\mathrm{L}}(q)  \\
\beta_{\mathrm{R}}(q)  
\end{pmatrix} = 
    \begin{pmatrix} 
r & t' \\
t & r' 
\end{pmatrix}
\begin{pmatrix} 
\alpha_{\mathrm{L}}(q)  \\
\alpha_{\mathrm{R}}(q)  
\end{pmatrix},
\end{equation}
where $r$ $(r')$ and $t$ $(t')$ are the reflection and transmission amplitudes, respectively, for a spin wave originating from the left (right) lead. 

In the left lead, we can express $a_j$ as~\footnote{An analogous expression can be used for the right lead.}
\begin{equation}
    a_{j} = \int_0^{\pi / d} \frac{dq}{2 \pi/d} \left[e^{ i q jd} \alpha_{\mathrm{L}}(q) + e^{ i q jd} \beta_{\mathrm{L}}(q)\right].
    \label{Eq:parameterization}
\end{equation} 
If we substitute Eq. \eqref{Eq:parameterization} and its complex conjugate into Eq. \eqref{Eq:CurrentOperator} and utilize that the leads are in thermal equilibrium with the reservoirs such that $\langle \alpha_{\mathrm{L,R}}^{\dagger}(q_1) \alpha_{\mathrm{L,R}}(q_2) \rangle = \frac{2 \pi}{d}  \delta(q_1-q_2) f_{\mathrm{L,R}}(q_1)$, we find that
\begin{equation}
    \langle I_{j,j+1} \rangle = \frac{1}{2 \pi} \int_{\epsilon_{\mathrm{min}}}^{\epsilon_{\mathrm{max}}} d \epsilon\, T(\epsilon) \left(f_{\mathrm{L}}(\epsilon) - f_{\mathrm{R}}(\epsilon) \right).
\end{equation}
In this expression, $f_{\mathrm{L,R}}(\epsilon)$ represents the Bose-Einstein distributions in the left and right reservoirs, respectively, and $T(\epsilon) = \vert t \vert^2$. The integration limits are obtained from Eq.~\eqref{eq:dispersion_relation}.

Assume that the spin accumulation in the left lead is $\mu_{\mathrm{L}} = \mu + \delta \mu$ and that the spin accumulation in the right lead is $\mu_{\mathrm{R}} = \mu$, where $\delta \mu/\mu \ll 1$. We find that in the linear response, the spin conductance is given by
\begin{equation}
    G = \frac{1}{2\pi} \int_{\tilde{\epsilon}_{\mathrm{min}}}^{\tilde{\epsilon}_{\mathrm{max}}} d \tilde{\epsilon}\, T(\tilde{\epsilon}) \left(- \frac{d f}{d \tilde{\epsilon}}\right).
    \label{Eq:SpinConductivity}
\end{equation}
This result can also be derived using Green's functions~\cite{PhysRevB.69.174403}.
In Eq. \eqref{Eq:SpinConductivity}, we are integrating over the dimensionless energies $\tilde{\epsilon} = \epsilon/J$. Energies outside of the integration interval result in the spin waves in Eqs. \eqref{Eq:leftAnsatz} and \eqref{Eq:rightAnsatz} becoming evanescent waves that do not contribute to the spin conductance.
%%%%%%%%%%%%%%%%%%%%%%%%%%%%%%%%%%%%%%%%%%%%%%%%%%
%%%%%%%%%%%%%%%%%%%%%%%%%%%%%%%%%%%%%%%%%%%%%%%%%%%
\section{Results and discussion}
\label{sec:results}
For each realization of the system, we find that $r_x = r_y \equiv r$ and that $t_x = t_y \equiv t$, reflecting the fact that inside the leads, the spin waves are circularly polarized. Furthermore, we define $R = \vert r \vert^2$ and $T = \vert t\vert^2$ as the reflection and transmission probabilities, respectively, and find that $R + T = 1$. Since $R$ and $T$ depend on the realization of the system, we must perform an ensemble average $\langle \dots \rangle$ to obtain physically meaningful quantities. In our calculations, we used $10^3$ different realizations for the random vectors $\mathbf{n}_i$. In Fig.~\ref{fig:avgT.png}, we plotted $\langle T\rangle$ and $\langle \ln T \rangle$ as a function of $\tilde{\epsilon}$ for different values of $K/J$ and a fixed system length $L = N_{\mathrm{rand}} d$. In the remainder of this paper we set $d = 1$ for convenience.

\subsection{Transmission probability}
\label{Sec:Transmission probability}
\begin{figure}[h] %This graph was made in averageT.m
            \centering
            \includegraphics[width=\linewidth]{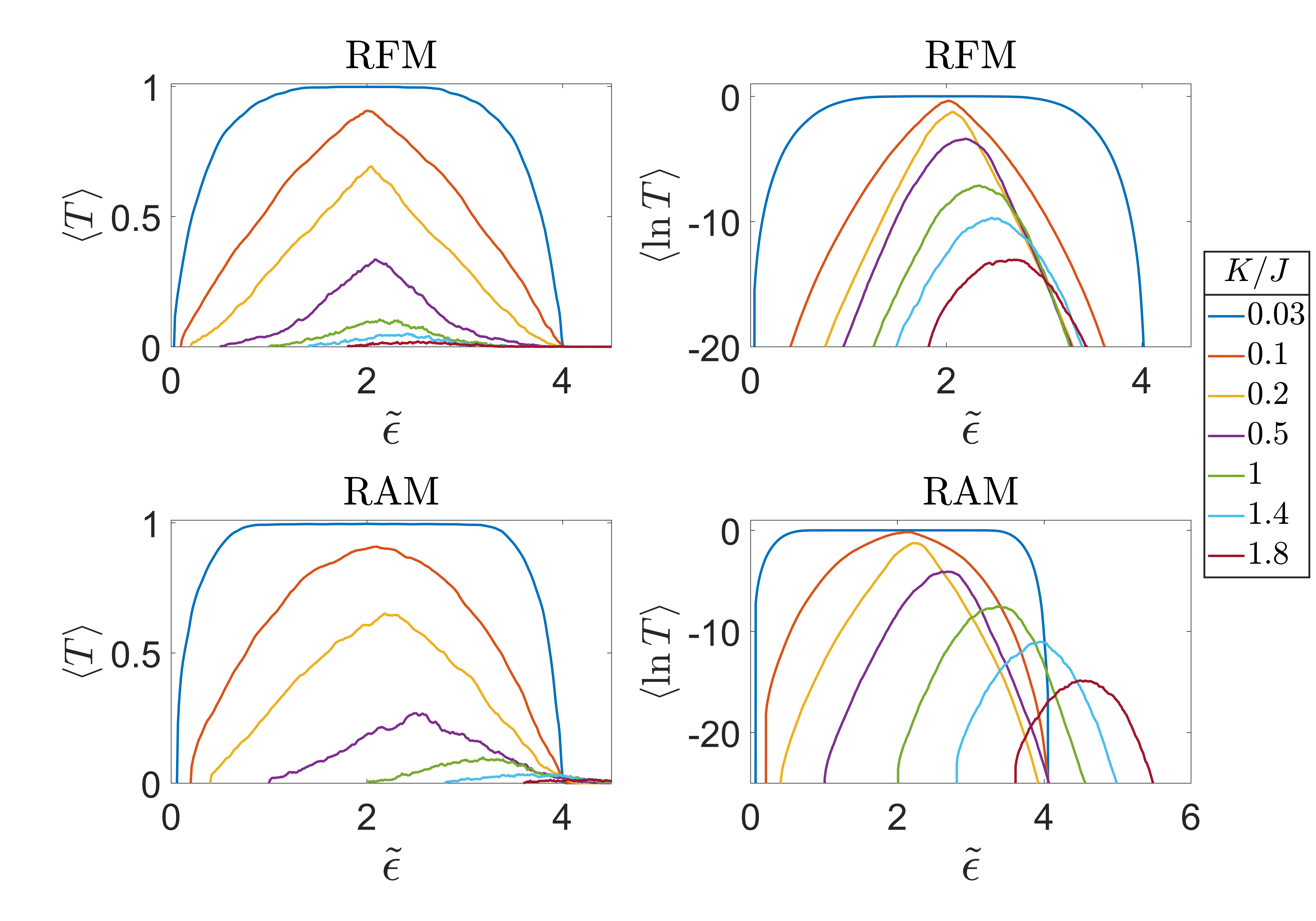}
            \caption[]%
            {{\small The behaviors of $\langle T \rangle$ and $\langle \ln T \rangle$ as a function of $\tilde{\epsilon}$ and $K/J$.  }}    
            \label{fig:avgT.png}
        \end{figure}

As the system becomes more disordered, the transmission probability decreases for both the RFM and the RAM. However, Fig.~\ref{fig:avgT.png} demonstrates that the quantitative behavior of the localization is significantly different in the two models. In both models, as $K/J$ increases, the maxima $\langle T \rangle_\mathrm{max}$ and $\langle \ln T \rangle_\mathrm{max}$ shift towards higher $\tilde{\epsilon}$, but in the RAM, this shift is greater than that in the RFM. In addition, the peak in the transmission probability is wider in the RAM compared to the RFM for small $K/J$. Thus, a broader range of spin waves can pass through the RAM compared to the RFM in the limit of weak disorder.
        
We can understand the difference in width from the Hamiltonian in Eq. \eqref{Eq:SpinHamiltonian}. In the RAM, the term causing disorder is $\left(\mathbf{S}_i \cdot \mathbf{n}_i\right)^2$; thus, the spin $\mathbf{S}_i$ wants to point either parallel or antiparallel to $\mathbf{n}_i$. The spin is also coupled to its neighbors through the exchange interaction. Therefore, in the RAM, whether the spin $\mathbf{S}_i$ chooses to point parallel or antiparallel to $\mathbf{n}_i$ depends on the neighboring spins. Meanwhile, in the RFM, the term causing disorder is $\mathbf{S}_i\cdot \mathbf{n}_i$, and the spin wants to only point parallel to $\mathbf{n}_i$. The ability to select whether to point parallel or antiparallel to $\mathbf{n}_i$ leads to the spin chains in the RAM being less disordered than the spin chains in the RFM, which in turn leads to a broader peak in the transmission probability.
%%%%%%%%%%%%%%%%%%%%%%%%%%%%%%%%%%%%%%%%%%%%%%%%%%%
\subsection{Self-averaging}
\label{Self-averaging quantities}
In disordered systems, certain quantities are not self-averaging in the thermodynamic limit. This is well known in disordered fermionic systems and is expected to be a general feature of a broad spectrum of disordered systems~\cite{PhysRevLett.77.3700}. A test to determine whether a quantity $O$ is self-averaging is to check whether the relative variance $\mathrm{RV}_O = \mathrm{Var}(O)/\langle O \rangle^2 $ vanishes (or is sufficiently small) in the limit $L\xrightarrow{} \infty$. For the fermionic 1D Anderson model with on-site disorder, one finds that the transmission probability and hence the conductance are not self-averaging~\cite{doi:10.1142/7663}. In two and three dimensions, one finds that the logarithms $\ln(T), \ln (G)$ are self-averaging such that $\mathrm{RV}_{\ln G} \sim L^{-D}\, (D = 2,3)$~\cite{PhysRevB.33.8441,PhysRevLett.77.3700}. In one dimension at finite temperature, one finds that $\ln G$ is only marginally self-averaging because $\mathrm{RV}_{\ln G}$ decays logarithmically with $L$~\cite{PhysRevLett.121.136806,PhysRevLett.53.2042,PhysRevB.33.8441}.

\begin{figure}[h] %Graph came from relativeVariance_kj04.m
            \centering
            \includegraphics[width=\linewidth]{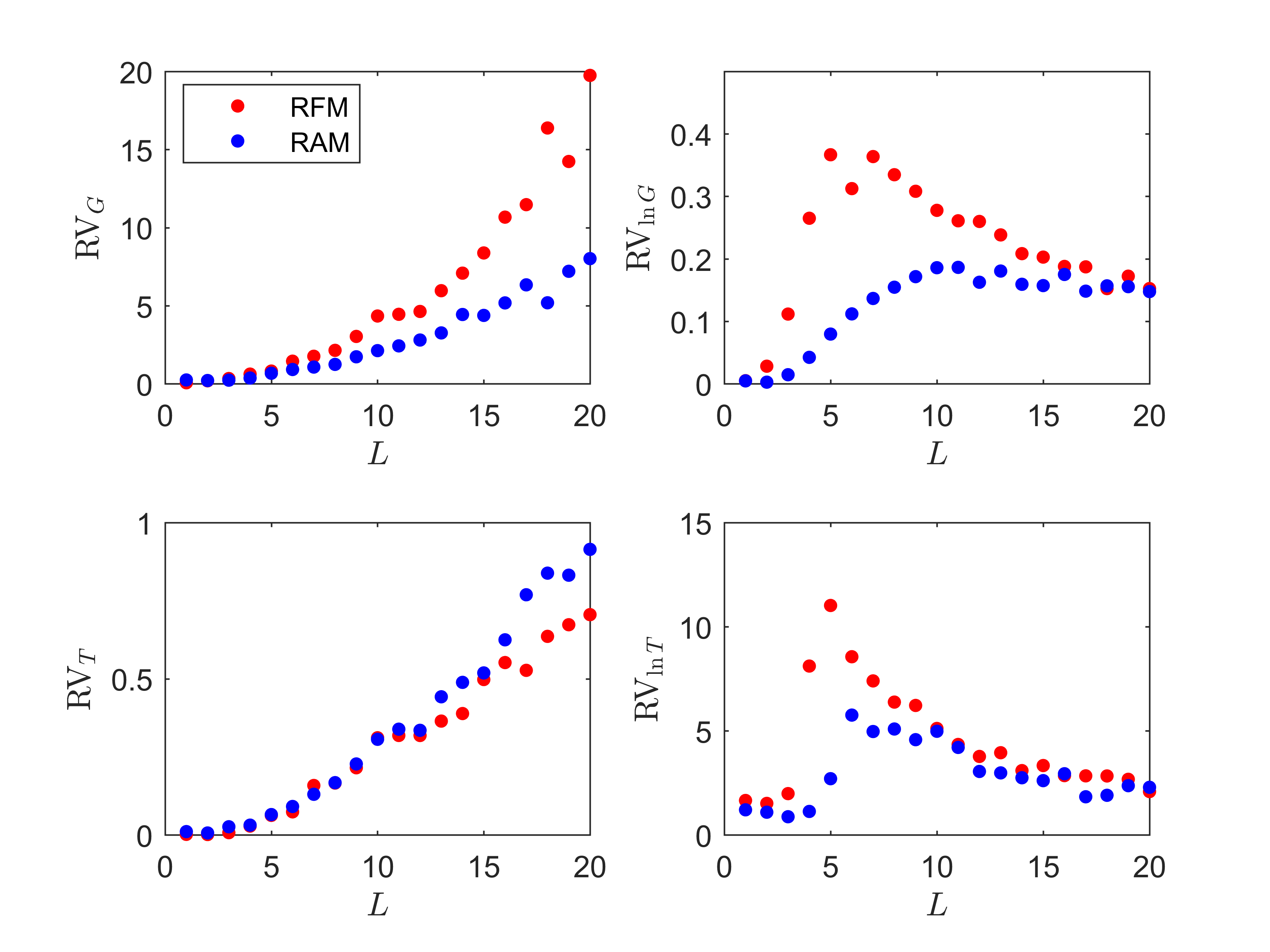}
            \caption[]%
            {{\small The length dependence of the relative variances $\mathrm{RV}_G$, $\mathrm{RV}_{\ln G}$, $\mathrm{RV}_T$, and $\mathrm{RV}_{\ln T}$ for the RFM and the RAM. The strength of disorder is $K/J = 0.4$.}}    
            \label{fig:relativeVariance.png}
        \end{figure}

As expected, we find similar results in this work. Fig.~\ref{fig:relativeVariance.png} shows that the relative variances $\mathrm{RV}_T$ and $\mathrm{RV}_G$ increase with the length of the system. In addition, the relative variances $\mathrm{RV}_{\ln T}$ and $\mathrm{RV}_{\ln G}$ decrease with the length of the system. Hence, as the length of the system increases, the fluctuations in $T$ and $G$ become much greater than the corresponding expectation values, meaning that they are not representative variables in the thermodynamic limit. Therefore, we use $\{\ln(T),\ln(G)\}$, rather than $\{T,G\}$, to calculate the localization lengths of the system.

\subsection{The localization length}
\label{The localization length}
In this work, it is natural to define two types of localization lengths. The first is based on the maximum of the transmission amplitude $\langle \ln T \rangle_\mathrm{max}$ in Fig.~\ref{fig:avgT.png}. The second is based on the conductance $\langle \ln G\rangle$. We refer to these localization lengths as $\tilde{L}_{\ln T}$ and $\tilde{L}_{\ln G}$, respectively.

\begin{figure}[h] %Graph came from relativeVariance_kj04.m
            \centering
            \includegraphics[width=\linewidth]{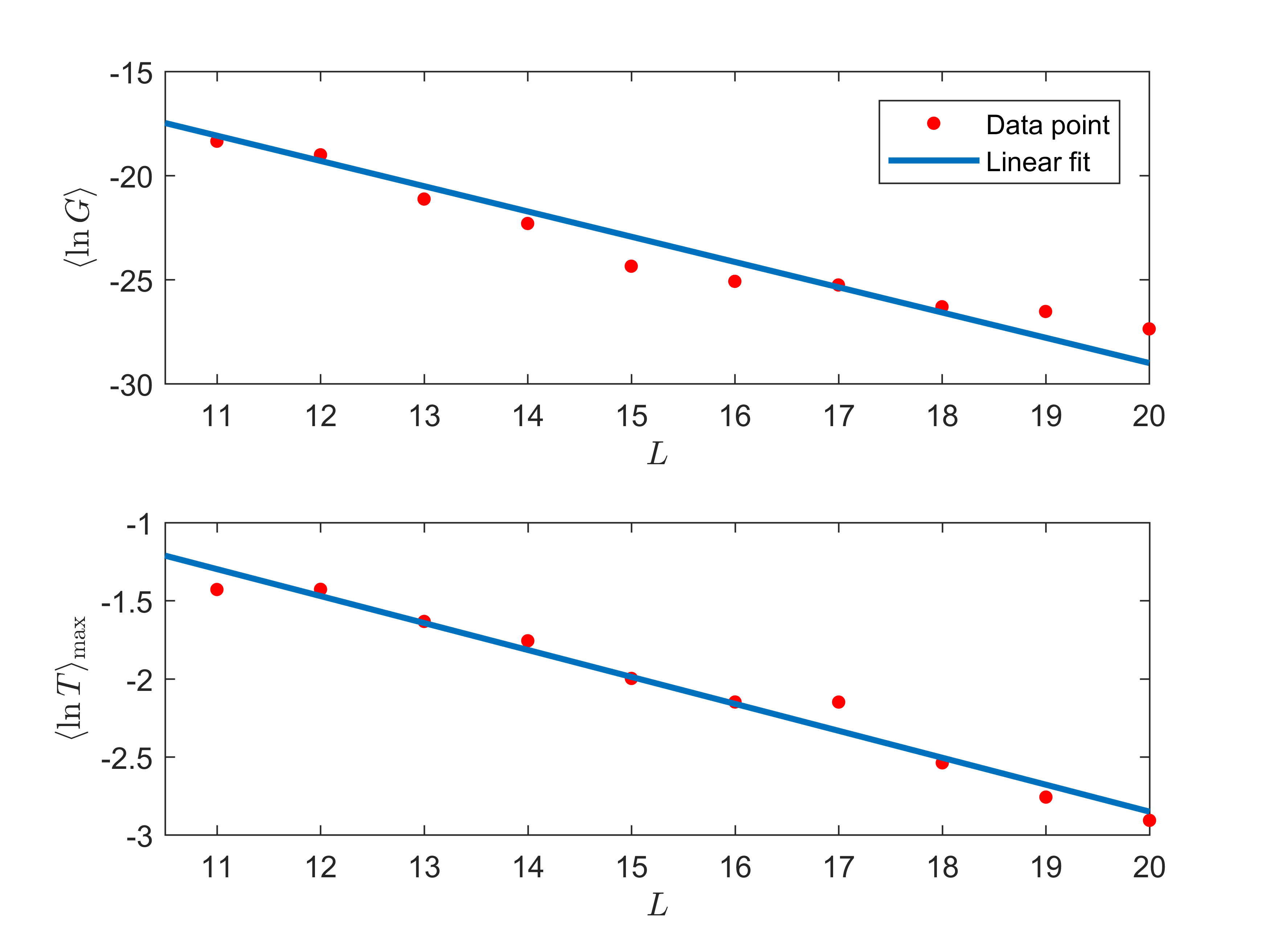}
            \caption[]%
            {{\small The length dependence of $\langle \ln G \rangle$ and $\langle \ln T \rangle_{\mathrm{max}}$ for the RFM can be approximated with a linear fit. The strength of disorder is $K/J = 0.4$, and the temperature is $\tilde{T} = 0.05$.}}    
            \label{fig:avgGandT.png}
\end{figure}

In Fig.~\ref{fig:avgGandT.png}, we plotted $ \langle\ln T\rangle_\mathrm{max} $ and $\langle \ln G\rangle$ as a function of the system length $L$ for a fixed $K/J$ and temperature $\tilde{T} = kT/J$. We have performed a curve fit with the functions
\begin{equation}
    \langle \ln T \rangle_{\mathrm{max}} = \frac{L}{\tilde{L}_{\ln T}} + A
    \label{Eq:linearCurveFitLogT}
\end{equation}
and
\begin{equation}
    \langle \ln G \rangle = \frac{L}{\tilde{L}_{\ln G}} + B
    \label{Eq:linearCurveFitLogG}
\end{equation}
such that the localization length can be extracted as the gradient of the straight lines in Fig.~\ref{fig:avgGandT.png}. In this particular case, we found a coefficient of determination $R^2$ with the value $R^2 = 0.95$ indicating a good fit.
To determine the localization lengths as a function of $K/J$, we performed straight line curve fits for graphs such as those found in Fig.~\ref{fig:avgGandT.png} but with different $K/J$ and $\tilde{T}$. In all cases, we found that the coefficient of determination was in the range $(0.9,1)$ and that the average coefficient was $\langle R^2 \rangle = 0.95$, indicating reasonably good straight-line fits. By then calculating the gradient of these straight lines, we can estimate the localization lengths as a function of $K/J$.

\subsubsection{Localization length from transmission}
\begin{figure}[h] %Graph came from localizationLengthFromLogTBothCurvesInSameFig
            \centering
            \includegraphics[width=\linewidth]{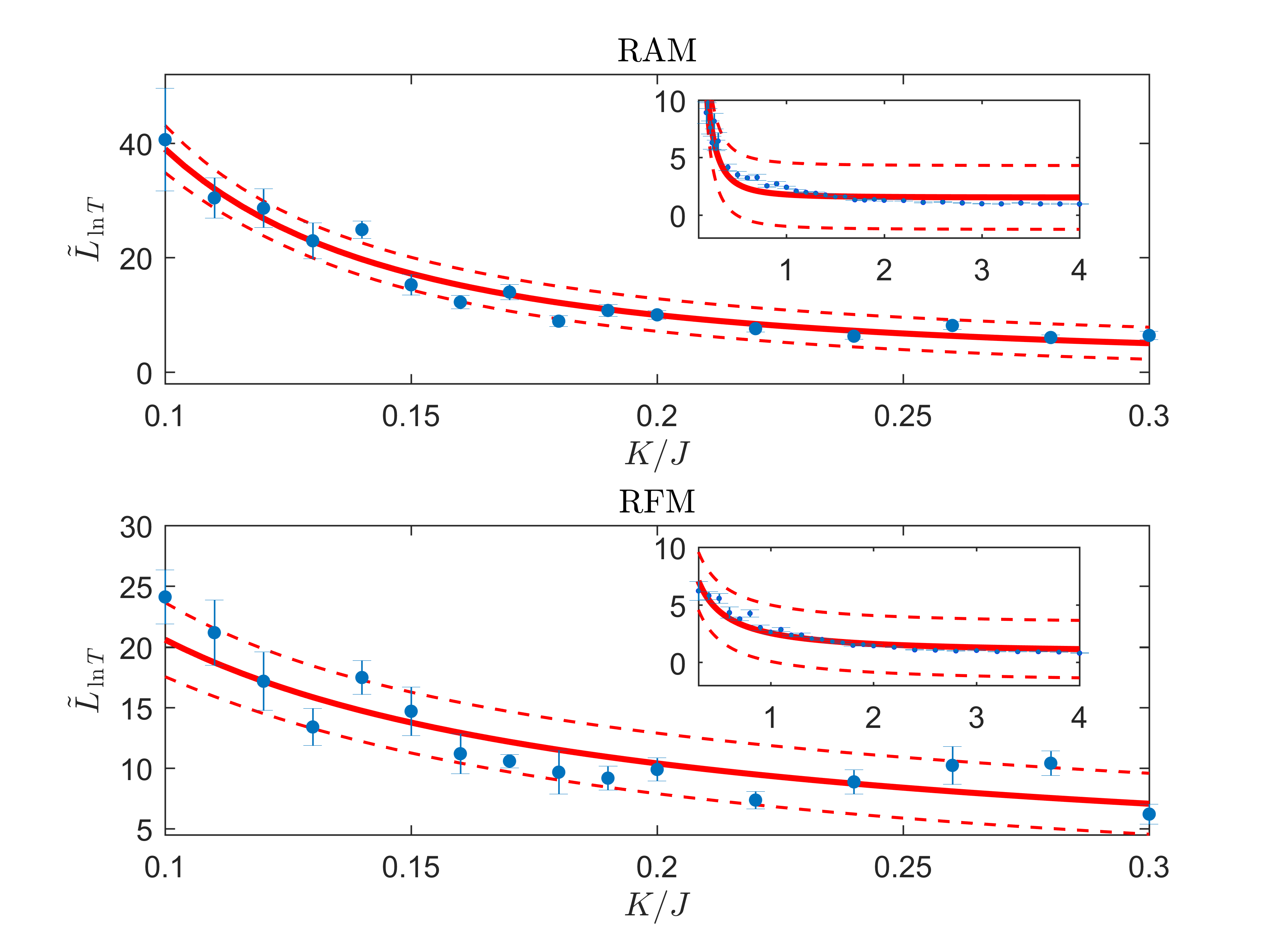}
            \caption[]%
            {{\small The behavior of $\tilde{L}_{\ln T}$ as a function of $K/J$ for the RAM and the RFM, respectively. The line represents the numerical fit in Eq. \eqref{Eq:LocLengthCurveFitFunction}, the dashed lines represent the $95\%$ confidence interval, and the points with error bars represent the localization length calculated from Eqs. \eqref{Eq:LLGx} and \eqref{Eq:LLGy} with standard error.}}    
            \label{fig:locLength_transmission}
        \end{figure}

In Fig.~\ref{fig:locLength_transmission}, we plot the localization length $\tilde{L}_{\ln T}$ and the $95\%-$confidence interval for the RAM and RFM, respectively. In both cases, we have performed a curve fit with the function
\begin{equation}
    \tilde{L}_{\ln T} = \eta\left(\frac{K}{J}\right)^{\nu} + \xi.
    \label{Eq:LocLengthCurveFitFunction}
\end{equation}
The parameters (with confidence intervals) are displayed in Table~\ref{tab:LocLengthParameterLnT}. Similar to fermionic systems~\cite{Billy2008}, we find that the localization length decays monotonically as a power law as we increase the strength of disorder. Our result can be made more universal by introducing the exchange length such that
\begin{equation}
    \tilde{L}_{\ln T} = \eta \left(l_{\mathrm{ex}}\right)^{-2\nu} + \xi.
\end{equation}

Note that for weak disorder, the localization length is greater in the RAM than in the RFM. This is a consequence of the fact that the spin chains are less disordered in the RAM compared to the RFM, as we discussed at the end of section~\ref{Sec:Transmission probability}.

\begin{table}[h]
	\centering
		\caption{The numerical values of the parameters in Eq. \eqref{Eq:LocLengthCurveFitFunction} for the RFM and the RAM. The brackets $(\dots)$ give the $95\%$ confidence interval.} 
	\begin{tabular}{l l l} \hline \hline
	 & RFM & RAM \\ \hline
    $\eta$ & $1.3\, (0.7,2.0)$ & $0.2\, (0.1,0.3)$ \\ \hline
    $\nu$ & $-1.2\, (-1.4,-1.0)$ & $-2.2\, (-2.4,-2.0)$\\ \hline
    $\xi$ & $1.1\, (0.3,1.9)$ & $1.6\, (1.0,2.1)$
    \\ 
    \hline \hline
	\end{tabular}
	\label{tab:LocLengthParameterLnT}
\end{table}

\subsubsection{Localization length from conductance}
\begin{figure}[h]%Made in locLengthConductivity.m
            \centering
            \includegraphics[width=\linewidth]{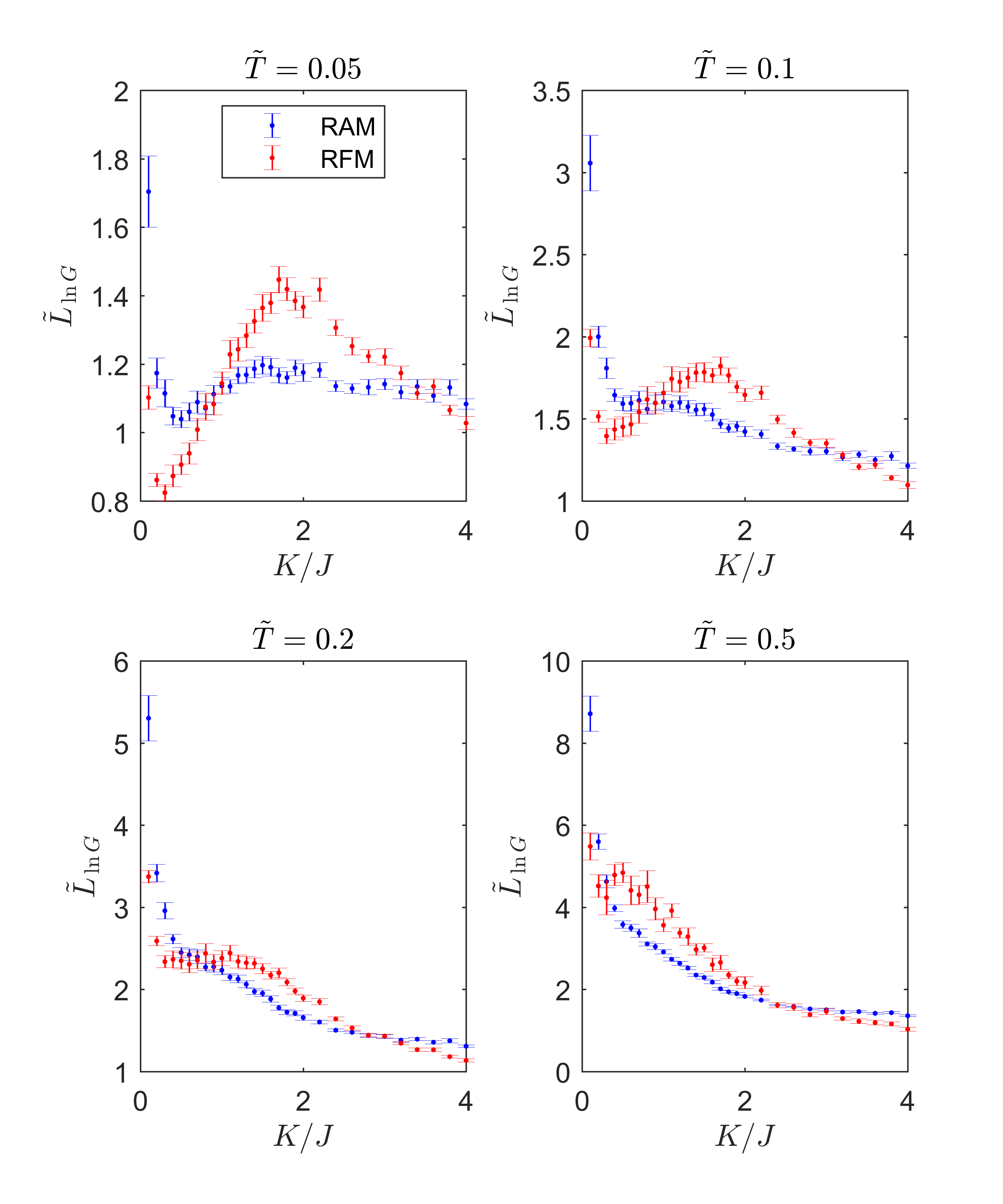}
            \caption[]%
            {{\small The temperature dependence of $\tilde{L}_{\ln G}$} for strongly disordered magnetic insulators.}    
            \label{fig:locLength_conductance}
\end{figure}

In Fig. \eqref{fig:locLength_conductance}, we plot $\tilde{L}_{\ln G}$ as a function of $K/J$ for different temperatures $\tilde{T}$. There is an interval $K/J \approx (0.5,2)$ where the localization length increases for small $\tilde{T}$. Furthermore, for sufficiently large $\tilde{T}$, this interval vanishes such that the localization length decays monotonically for all $K/J$. This nontrivial behavior arises because there is a competition between the temperature dependence of the broadening function $-df/d\tilde{\epsilon}$ and the disorder dependence of the transmission probability $T(\tilde{\epsilon})$ in Eq. \eqref{Eq:SpinConductivity}. As the temperature increases, the broadening function excites an increasing number of magnons, which in turn leads to a greater conductance. Meanwhile, as the system becomes more disordered, the transmission probability $T(\tilde{\epsilon)}$ decreases, resulting in a smaller conductance. On the interval $K/J \approx (0.5,2)$, the increase in conductance due to temperature is greater than the decrease in conductance due to disorder, which results in an increase in localization length.
Furthermore, in this interval, the localization length is comparable to the lattice spacing $d$, which means that there may be complicated microscopic details of the model that may further enhance this effect.

\begin{figure}[h] %This graph was made in localization.m
            \centering
            \includegraphics[width=\linewidth]{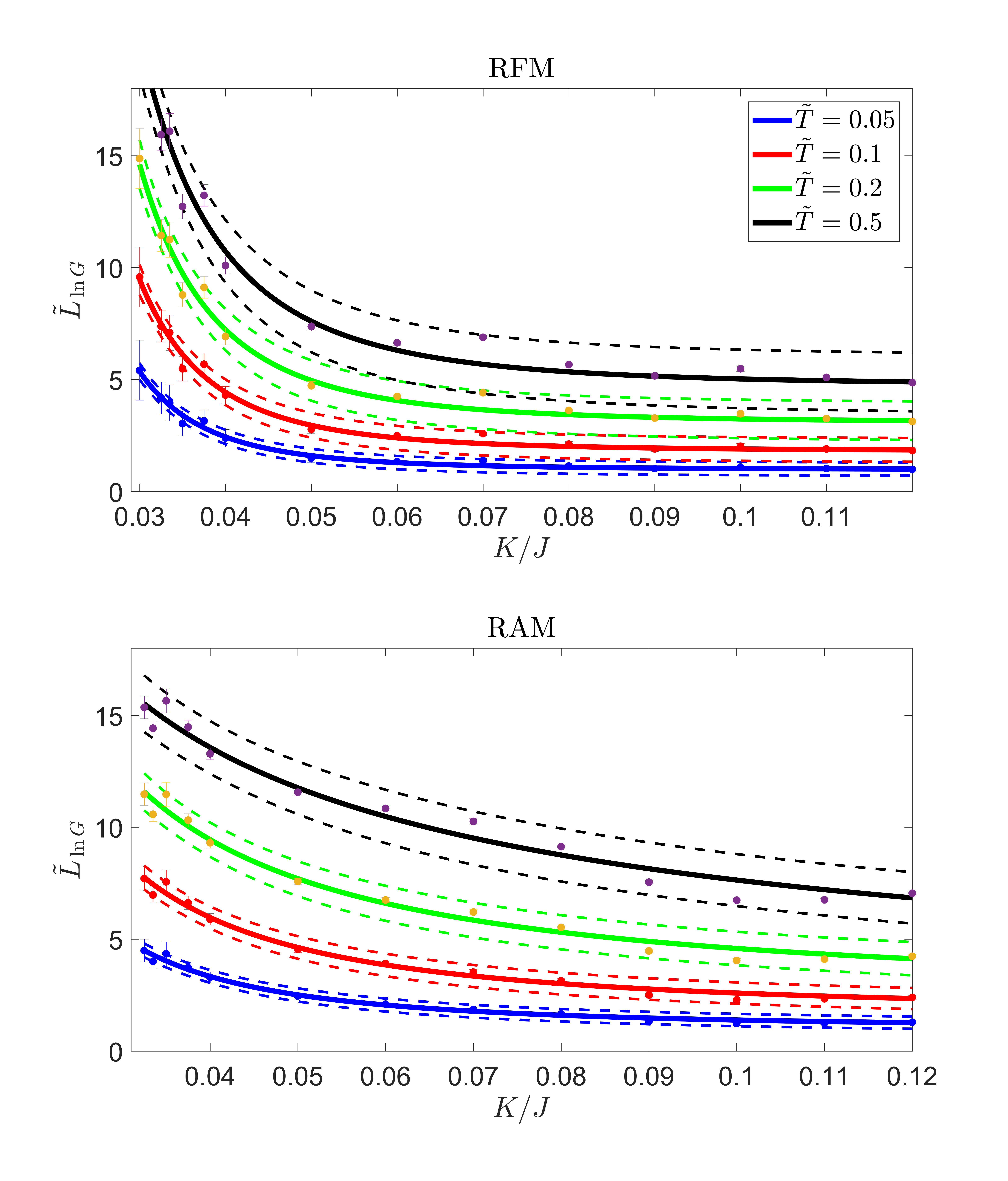}
            \caption[]%
            {{\small The temperature dependence of $\tilde{L}_{\ln G}$} in the limit of weak disorder $K/J \xrightarrow{} 0.$ }    
            \label{fig:smallLocalizationLengthFromG}
        \end{figure}

Due to the complicated temperature and disorder dependence, it is numerically challenging to determine a closed formula such as the one in Eq. \eqref{Eq:LocLengthCurveFitFunction} for the localization length $\tilde{L}_{\ln G}$. However, in the weak-disorder limit $K/J \xrightarrow{} 0$, it is reasonable to assume that the localization length decays as a power law of the form
\begin{equation}
L_{\ln G} \sim \left(\frac{K}{J}\right)^{\gamma} = \left(l_{\mathrm{ex}}\right)^{-2\gamma}
\label{Eq:CriticalExponentConductance}
\end{equation}
where $\gamma$ is the critical exponent. Fig.~\ref{fig:smallLocalizationLengthFromG} shows the result of such a curve fit for the RFM and the RAM for different temperatures. The corresponding critical exponents $\gamma_{\mathrm{RFM}}$ and $\gamma_{\mathrm{RAM}}$ are given in Table~\ref{tab:LocLengthParameterLnG}.

\begin{table}[h]
	\centering
		\caption{The numerical values of the critical exponent for the RFM and the RAM for different temperatures. The brackets $(\dots)$ provide the $95\%$ confidence interval.} 
	\begin{tabular}{l l l} \hline \hline
	$\Tilde{T}$ & $\gamma_{\mathrm{RFM}}$ & $\gamma_{\mathrm{RAM}}$ \\ \hline
   0.05 & $-3.8\,(-4.3,-3.4)$ & $-1.9\,(-2.3,-1.6)$\\ \hline
   0.1 & $-3.7\,(-4.2, -3.2)$ & $-1.7\,(-2.0,-1.4)$ \\ \hline
   0.2 & $-3.6\,(-4.1,-3.1)$ & $-1.3\,(-1.6,-1.0)$ \\ \hline
   0.5 & $-3.3\,(-3.8,-2.8)$ & $-0.7\,(-1.0,-0.4)$ \\ \hline
    \hline
	\end{tabular}
	\label{tab:LocLengthParameterLnG}
\end{table}

In our simulations, we kept the temperature below the Curie temperature $\tilde{T} = 1$, where the temperature fluctuations of the spins are negligible. For temperatures $\tilde{T} \approx 1$, there will be additional temperature-induced disorder. This issue has previously been investigated~\cite{PhysRevB.89.024409,EveresNowak:disorder2015,PhysRevB.97.214415} by including a temperature-dependent stochastic field in the effective field $\mathcal{H}_i^{\kappa}$ in Eq. \eqref{Eq:LLG}, and it was found that temperature fluctuations shorten the localization length and enhance the Anderson localization.

%%%%%%%%%%%%%%%%%%%%%%%%%%%%%%%%%%%%%%%%%%%%%%%%%%%
\section{Summary and Conclusions}
\label{sec:conclusions}

In this paper, we have applied the Landauer-B{\"u}ttiker formalism to noncollinear disordered magnetic insulators. We have considered both amorphous magnets and magnets with spin disorder modeled by the RAM and the RFM respectively. We calculated the self-averaging quantities $\langle \ln T \rangle$ and $\langle \ln G \rangle$ as a function of system length $L$ for a broad range of disorder strengths $K/J$. Consistent with the literature, we found evidence for Anderson localization such that $\langle \ln T \rangle$ and $\langle \ln G \rangle$ were linear functions of the system length $L$. This allowed us to define two localization lengths $\tilde{L}_{\ln G}$ and $\tilde{L}_{\ln T}$ based on the conductance and the maximum transmission probability, respectively. In the limit of weak disorder, the localization lengths obeyed power laws, and we calculated the relevant critical exponents. We expect our results to be general because they are expressed through the universal exchange length $l_{\mathrm{ex}}$.

We found that the Anderson localization is more prominent in the RFM than in the RAM. The reason for this result is that the competition between the exchange interaction and the disorder term leads to more disordered spin chains in the RFM than in the RAM. The spin chains in the RAM are less disordered because the disorder arises from a random anisotropy, where the spin can point either parallel or antiparallel to the anisotropy with the same energy cost. Whether the spin points parallel or antiparallel to the anisotropy is determined by the neighboring spins through the exchange interaction, and consequently, the configuration with the least disorder will be chosen by the system.

The results obtained here are valid in the limit of quenched disorder, i.e., $\tilde{T}\ll 1$, where the random field and anisotropy are temperature independent. If the temperature is close to the Curie temperature of the system, one must include temperature fluctuations in the Landau-Lifshitz equations. Such effects have been considered in other works, and it has been shown that temperature fluctuations decrease the localization length.

To experimentally verify the critical exponents obtained in this paper, we propose a setup in which a disordered magnetic nanowire is sandwiched between two normal metals. Similar setups for ordered magnets have been considered in other works~\cite{PhysRev.95.1154,DYAKONOV1971459,PhysRevLett.85.393,Murakami1348,PhysRevLett.92.126603,Jungwirth2012,Kajiwara2010,doi:10.1063/1.2199473,Valenzuela2006,PhysRevLett.98.156601,PhysRevLett.99.266603,Seki2008,PhysRevB.78.014413,doi:10.1063/1.3327809,Cornelissen2015}. By applying a charge current in the left metal, the spin Hall effect generates a spin current through the disordered nanowire and into the right metal. This will give rise to a spin wave propagating through the hybrid structure and into the right metal, where the spin current is converted into a charge current via the inverse spin Hall effect.

Alternatively, we can instead sandwich a disordered magnet between two ferromagnetic leads. We can excite spin waves in the left ferromagnet by applying a microwave with the ferromagnet resonance frequency. This spin wave will then propagate through the disordered insulator and into the right ferromagnet, where the resulting spin current can be measured. By measuring the spin current propagating through the disordered region, one should be able to characterize the localization length in terms of the critical exponents of the system.

%%%%%%%%%%%%%%%%%%%%%%%%%%%%%%%%%%%%%%%%%%%%%%%%%%%
\acknowledgments
The authors would like to express their thanks to R. A. Duine for fruitful discussions of the subject.

This research was supported by the European
Research Council via Advanced Grant No. 669442
''Insulatronics'' and the Research Council of Norway through
its Centres of Excellence funding scheme, Project
No. $262633$, ''QuSpin''.

%%%%%%%%%%%%%%%%%%%%%%%%%%%%%%%%%%%%%%%%%%%%%%%%%%%

%%%%%%%%%%%%%%%%%%%%%%%%%%%%%%%%%%%%%%%%%%%%%%%%%%%

%%%%%%%%%%%%%%%%%%%%%%%%%%%%%%%%%%%%%%%%%%%%%%%%%%%

%%%%%%%%%%%%%%%%%%%%%%%%%%%%%%%%%%%%%%%%%%%%%%%%%%%

%\bibliography{references}


%%%%%%%%%%%%%%%%%%%%%%%%%%%%%%%%%%%%%%%%%%%%%%%%%%%

\end{document}